\documentclass[a4paper]{jpconf}
\usepackage{graphicx}
\begin{document}
\title{Problems and Progress in Covariant High Spin Description}

\author{Mariana Kirchbach and V{\'{i}}ctor Miguel Banda Guzm\'an}

\address{Instituto de Fis{\'{i}}ca, Universidad Aut\'onoma de  San Luis Potos{\'{i}}, Av. Manuel Nava 6, Zona Universitaria, San Luis Potos{\'{i}}, SLP 78290,
M\'exico}

\ead{mariana@ifisica.uaslp.mx, vmbg@ifisica.uaslp.mx}

\begin{abstract}
A universal description of particles with spins $j\geq 1$, transforming in  $(j,0)\oplus (0,j)$, is developed by means of representation specific second order differential wave equations without auxiliary conditions and in covariant bases such as  Lorentz tensors for bosons, Lorentz-tensors with Dirac spinor components for fermions, or, within the basis of  the more fundamental Weyl-Van-der-Waerden $sl(2,C)$ spinor-tensors. At the root of the method, which is free from the pathologies suffered by the traditional approaches, are projectors constructed from the Casimir invariants of the spin-Lorentz group, and the group of translations in the Minkowski space time. 
\end{abstract}

\section{Introduction}
High-spin $j\geq 1$ fields, both massive and massless, always have counted to the main topics in field theories. In particle physics such fields describe hadron resonances whose spins vary from $1/2$ to $17/2$ for baryons, and from zero to six for mesons \cite{PART}. 
At hadron colliders, intermediate high spin resonances  can be produced which subsequently decay in  a pair of photons, a process of high  interest in the literature \cite{DiphotonPhen}. In gravity, high-spin bosons can couple to the metric tensor and cause its deformation \cite{Gravity}, and are besides this fundamental to the physics of rotating black holes \cite{rot_BH}. Also gravitational interactions between high-spin fermions are of interest \cite{A}. 
The traditional  approaches to  high-spin fields have been developed between 1939 and 1964  (see \cite{HSReview} for a recent review, and
\cite{BR} for a standard textbook) by  Fierz and Pauli (FP) \cite{PF}, Rarita and Schwinger \cite{RS},  Bargmann and Wigner (BW) \cite{BW1948},
Joos \cite{Joos} and Weinberg \cite{Weinberg}. 
In the next section we briefly highlight these four methods, comment on their problems, and report in section 3 on a new approach, suggested in \cite{We}-\cite{BWpost}, before closing with a brief summary.

\section{The traditional methods for high-spin description and their problems}\label{sec1} 

The methods for high-spin description are distinct through the type of the representation spaces (reps) used to embed the spin of interest.
Below we list  various possibilities to describe spin-$j$.

\subsection{ Multiple-spin valued representation spaces: the totally symmetric Lorentz  tensors for bosons, and Lorentz--tensor--Dirac--spinors for fermions. The Fierz-Pauli and Rarita-Schwinger frameworks}
\vspace{0.1cm}
Particles of spins, $j>\frac{1}{2}$, have been so far frequently 
described in terms of $so(1,3)$ representation spaces of  multiple spins and alternating parities, corresponding to the  totally symmetric $\left(\mbox{\bf Sym} \right)$ Lorentz tensors of rank-$j$,   
\begin{equation} 
\mbox{\bf Sym}\, \Phi_{\mu_1...\mu_j}\simeq \left(\frac{j}{2},\frac{j}{2}\right), 
\end{equation} 
for bosons (Fierz-Pauli framework \cite{PF}), or, the totally symmetric rank-$(j-1/2)$ Lorentz tensors with Dirac spinor components, $\psi$, 
\begin{equation} 
\mbox{\bf Sym}\, \psi_{\mu_1...\mu_{j-\frac{1}{2}}}\simeq \left(\frac{j-\frac{1}{2}}{2}, 
\frac{j-\frac{1}{2}}{2}\right)\otimes \left[ 
\left(\frac{1}{2},0\right)\oplus \left(0,\frac{1}{2}\right)\right], 
\end{equation} 
for fermions (Rarita-Schwinger (RS) framework \cite{RS}), respectively. 
The particles have been associated with the degrees of freedom (d.o.f.) of the highest spins in the spaces under discussion, while the remaining degrees of freedom, corresponding to lower spins, had to be  projected out by properly chosen auxiliary conditions for the sake of 
ensuring the correct number of physical d.o.f. required by spin-$j$.
The  wave equations in the Fierz-Pauli approach to high-spin- $j$ bosons read,
\begin{eqnarray}
(\partial^2+m^2)\, \mbox{\bf Sym}\, \Phi_{\mu_1...\mu_i...\mu_{j}}=0, &\quad& \mbox{tr}\, \mbox{\bf Sym}\, \Phi_{\mu_1...\mu_i...\mu_{j}}=0, \quad \partial^{\mu_i}\mbox{\bf Sym}\, \Phi_{\mu_1...\mu_i...\mu_j}=0,
\end{eqnarray}
while the Rarita-Schwinger approach to spin-$j$ fermions is cast as, 
\begin{eqnarray} 
(i\partial\!\!\!/ -m) 
\mbox{\bf Sym}\, \psi_{\mu_1...\mu_i...\mu_{j-\frac{1}{2}}}=0,\,\,
\gamma^{\mu_i}\mbox{\bf Sym} \, \psi_{\mu_1...\mu_i...\mu_{j-\frac{1}{2}}}=0, && 
\partial^{\mu_i}\mbox{\bf Sym}\, \psi_{\mu_1...\mu_i...\mu_{j-\frac{1}{2}}}=0.
\end{eqnarray}

\subsection{ Single-spin valued   $(j,0)\oplus(0,j)$ ``bi-vector'' representation spaces. The Joos-Weinberg framework}
\vspace{0.1cm}
Alternatively, it has been considered that spin-$j$ particles can also be described  in terms of  single-spin 
valued  representation spaces of the even number of $2(2j+1)$ components,
\begin{eqnarray} 
\psi^{(j)}_{B} &\simeq & (j,0)\oplus (0,j),\quad 
B \in \left[1, 2(2j+1) \right],
\label{WJ-spinors} 
\end{eqnarray}
 a scheme  known as the  Joos-Weinberg framework \cite{Joos}, \cite{Weinberg}.
Differently from the RS approach, the $2(2j+1)$-component wave function, {\bf  column}$\left(\psi^{(j)}_1...\psi^{(j)}_{2j+1},\psi^{(j)}_{(2j+1)+1}...\psi^{(j)}_{2(2j+1)}\right) $, occasionally termed to as ``bi-vector'', satisfies one sole differential equation, which is  however of a high-order,
\begin{eqnarray} 
\left(i^{2j}\left[ 
\gamma_{\mu_1\mu_2...\mu_{2j}}\right]_{AB}\partial^{\mu_1}\partial^{\mu_2}... 
\partial^{\mu_{2j}}-m^{2j}\delta_{AB}\right)\psi^{(j)}_B(x)&=&0. 
\label{WJ-eqs} 
\end{eqnarray} 
Here, $\left[ \gamma_{\mu_1\mu_2...\mu_{2j}}\right]_{AB}$ are 
the elements of the 
generalized Dirac Hermitean matrices of dimensionality 
$\left[2(2j+1)\right]\times \left[2(2j+1)\right]$, which 
transform as Lorentz tensors of rank-$2j$. 
The complete sets of such matrices have been extensively studied in the 
literature 
for the purpose of constructing all the possible field bi-linears needed in the 
definitions of the 
generalized  currents, both  transitional  and diagonal \cite{Sankar}. 
Though well elaborated, the Weinberg-Joos formalism has attracted comparatively less attention than the linear Rarita-Schwinger framework
not only because of the difficult to  tackle high order of the differential equations and the high dimensionality of the generalized Dirac matrices. 
In addition, bi-vectors as a rule can not be equipped neither with Lorentz--, nor with Dirac indexes, a reason for which
their couplings to spinorial, vectorial etc targets have to be  described through  uncomfortable rectangular matrices.

\subsection{ Single-spin valued totally symmetric Weyl--Van--der--Waerden tensor-spinor representation spaces. The Bargmann-Wigner framework}
\vspace{0.1cm}
Another option for spin-$j$ description is provided by the totally symmetric product,
of $n=2j$ copies of a Dirac spinor, $\psi\simeq (1/2,0)\oplus (0,1/2)$, where $(1/2,0)$, and $(0,1/2)$ are the right-- and left-handed  Weyl--Van--der-Waerden two-component spinors corresponding to the two nonequivalent fundamental $sl(2,C)$ representation spaces, also  known as ``spinor'' ($\xi$), and ``co-spinor'' ($\dot\eta)$,
\begin{equation}
\mbox{\bf Sym}\, \Psi^{(n)}_{b_1...b_n}=\mbox{\bf Sym}\, \psi_{b_1}...\psi_{b_n}\simeq\mbox{\bf Sym}\, \left[ \left(\frac{1}{2}, 0 \right)\oplus \left(0,\frac{1}{2} \right)\right]_1 \otimes ...\otimes
\left[ \left(\frac{1}{2}, 0 \right)\oplus \left(0,\frac{1}{2} \right)\right]_n,
\end{equation}
where  $ b_i=1,2,3,4$, a scheme known as the Bargmann-Wigner method \cite{BW1948}.
The Bargmann-Wigner rank-$n$ spinor satisfies also a high-order differential equation which reads,
\begin{equation}
\left(\gamma_\mu p^\mu -m \right)^{a_1 b_1}...\left(\gamma^\mu p^\mu -m \right)^{a_i b_i}....\left(\gamma_\mu p^\mu -m \right)^{a_n b_n} \mbox{\bf Sym}\, \Psi_{b_1...b_i...b_n}=0.
\end{equation}
All four schemes are known to be plagued by serious inconsistencies, which are however  of distinct kinds. The  Fierz-Pauli and the Rarita-Schwinger fields have multiple redundant components which are eliminated by the auxiliary conditions solely at the free-particle level. Upon  couplings 
to external fields, the auxiliary conditions no longer serve their purpose, in consequence of which the wave fronts of the (classical) coupled solutions can propagate acausally (Velo-Zwanziger problem \cite{VeloZw}). A further inconsistency has been revealed by Johnson and Sudarshan \cite{SJ} who showed that the equal time anti-commutators between spin-$3/2$  fields, minimally coupled to the electromagnetic field,  are not positive definite. In addition, the attempts to construct Lagrangians within these methods are becoming increasingly involved due to difficulties of incorporating the auxiliary conditions at the interaction level, a reason for which
it is common to pursue alternative, though not fully covariant methods in data evaluations. Specifically in hadron collisions, where the  calculations of the cross sections are carried out  within the preferred  Center of Mass frame, it is common to account for the contributions of intermediate states of high-spin-$J$ to processes as, say,  diphoton production, by parametrizing the corresponding polarized decay amplitude by a single constant and factorizing a rotational $SO(3)$  Wigner function, $d^J_{m, \lambda -\lambda^\prime}(\theta)$, that encodes the angular dependence of the amplitude. Specifically in 
\cite{DiphotonPhen}, the absence of high-spin Lagrangians has been compensated by suggesting effective Lagrangians that may be used to simulate some particular intermediate resonances of both low and high spins, through Monte Carlo generators.\\

\noindent
On the other side, the Joos-Weinberg and the Bargmann-Wigner frameworks refer to a single wave equation without any need for auxiliary conditions, but they are of the order twice the particle's spin. Differential equations of the order higher than 2 are as a rule difficult to tackle because they lead to higher-order field theories which suffer severe inconsistencies such as ghost solutions of the bad type,  kinetic terms of  wrong signs,  states of negative norms, violation of unitarity etc.
Furthermore, particles in such theories can propagate non-locally. All the  experimentally verified fundamental  theories in modern physics are based on Lagrangians of second order in the momenta. Only in effective theories with over integrated fields, high order Lagrangians may appear. One of the reasons behind the problems caused by high order differential equations is the so  Ostrogradskian instability \cite{Woodard} which, at the level of, say, classical mechanics, for concreteness,  predicts phase spaces of unstable orbits, and needs a special effort to be handled \cite{TaiJunChen}.
In the light of the above discussion, it is desirable to search for a method which would allow for a consistent description of  any spin,
\begin{itemize}
\item  by one sole wave equation of maximally second order in the derivatives, and derivable from a Lagrangian,
\item  within  Lorentz-, or, Weyl-Van-der-Waerden tensor-spinor bases.
\end{itemize}
Such a formalism  for spin-$j$ in single-spin-- $(j,0)\oplus (0,j)$ reps has been designed in refs.~\cite{We},\cite{IJMPE}, \cite{BWpost}, and is briefly highlighted below.
\section{Progress in describing any spin-$j$ in $(j,0)\oplus (0,j)$ in Lorentz-tensor--, Lorentz-tensor-Dirac-spinor--, or, Weyl-Van-der-Waerden tensor-spinor bases and wave equations of second order.  The spin-Lorentz group projector method }
\vspace{0.1cm}
The first goal of the references \cite{We}, \cite{IJMPE}, \cite{BWpost}  has been to describe the pure spin $(j,0)\oplus (0,j)$ states, be it through Lorentz--, or through  Weyl-Van-der-Waerden spinor-tensors, this  for the sake of constructing by simple index contractions vertexes which involve interactions of high-spins with gauge fields, such as the photon, and/or  spinorial targets, such as the proton, and  thus avoid the introduction of the cumbersome  index-matching  rectangular matrices, typical  for the Joos-Weinberg formalism. In order to illustrate the essentials of the method, we here bring as representative examples the two simplest cases, beginning with the description of the  $(1,0)\oplus (0,1)$ field as a totally anti-symmetric Lorentz tensor of second rank, $B_{\left[ \mu,\nu\right]}$, with the brackets denoting index anti-symmetrization. On these grounds, in a next step, pure spin-$3/2$ can be described by means of that very totally-antisymmetric second-rank Lorentz tensor but now with Dirac spinor, $\psi\simeq (1/2,0)\oplus (0,1/2)$,  components, i.e. by the direct product,
$B\otimes \psi$. Indeed, within the latter representation space, a reducible one, one finds the irreducible $(3/2,0)\oplus (0,3/2)$ sector contained  as,
\begin{eqnarray}
B\otimes \psi
\longrightarrow\left[ \left(\frac{3}{2},0 \right) \oplus \left(0,\frac{3}{2} \right)\right]\oplus \left[ \left(\frac{1}{2},1 \right) \oplus \left(1,\frac{1}{2} \right)\right]&\oplus& \left[ \left(\frac{1}{2},0 \right)\oplus \left(0,\frac{1}{2} \right)\right].
\label{prdct_spcs}
\end{eqnarray}  
The latter equation shows that pure spin-$3/2$ transforming in $(3/2,0)\oplus (0,3/2)$ can be described in terms of a 24 component wave function, denoted by   $\left[\Psi^{(3/2,0)\oplus (0,3/2)}_{\left[ \mu,\nu \right]}\right]_a$, of two anti-symmetric 
Lorentz indexes, $\left[ \mu, \nu\right]$, and a Dirac index, $"a"$, taking the values, $a=1,2,3,4$,
 provided, one would be able to eliminate  from them the redundant $16$ components belonging to the $so(1,3)$ irreducible sectors,
$\left( \frac{1}{2},1 \right) \oplus \left(1,\frac{1}{2} \right)$, and $\left( \frac{1}{2},0 \right)\oplus \left(0,\frac{1}{2}\right)$. 
 Such an  elimination can indeed be realized by the aid  of momentum independent projectors constructed from the invariants of the spin-Lorentz group, a result due to \cite{We},\cite{IJMPE}.

\subsection{The spin-Lorentz group and  static covariant projectors on irreducible representation spaces}
\vspace{0.1cm}
The Lorentz group transforming the internal spin degrees of freedom, henceforth termed to as  spin-Lorentz group, and denoted by, ${\mathcal L}$,  is a subgroup of the complete Lorentz group, which  acts besides on the spin- also  on the external space-time degrees of freedom.
The ${\mathcal L}$ generators, denoted by $S_{\mu\nu}$, are quadratic $d\times d$ constant matrices, where $d$ fixes the finite dimensionality of the internal representation space, and encodes the spin value. For the special case of a pure spin, dimensionality and spin are related as
$d=2(2j+1)$, while for representations of multiple spins, relations of the type  $d=\sum_i(2j_i+1)$, or, $ d=2\sum_i (2j_i+1)$, can hold valid.
The algebra of the spin-Lorentz group, termed to as {\tt homogeneous spin-Lorenz group} (HSLG) reads \cite{KimNoz},
\begin{eqnarray}
{\mathcal L}:\quad \left[S_{\mu\nu},S_{\rho\sigma}\right] &=&i(g_{\mu\rho}S_{\nu \sigma}-g_{\nu\rho}S_{\mu\sigma}+g_{\mu\sigma}S_{\rho\nu}-g_{\nu\sigma}S_{\rho\mu}).
\label{Lrntz_algbr}
\end{eqnarray}
It has two Casimir invariants,
here denoted in their turn by, $F$ and $G$, and  defined as,
\begin{eqnarray}
F_{AB}= \frac{1}{4}\left[S^{\mu\nu}\right]_{AD}\left[S_{\mu\nu}\right]_{DB}, &&
G_{AB}=\frac{1}{8}\epsilon_{\mu\nu}\,\, ^{\alpha\beta}\left[S^{\mu\nu}\right]_{AC}\left[S_{\alpha\beta}\right]_{CB}, \,\,A,B, C, D,...=1,...,d.
\label{Cas2}
\end{eqnarray}
The two operators in  (\ref{Cas2}) identify  unambiguously any \underline{irreducible}
finite dimensional ${\mathcal L}$ group representation space, here generically denoted by, 
$\psi_{(j_1,j_2)\oplus (j_2,j_1)}=\phi^R_{(j_1,j_2)}\oplus \phi^L_{(j_2,j_1)}$, where $\phi^R_{(j_1,j_2)}$ and $\phi_{(j_2,j_1)}^L$ are in their turn its  
left- and right-handed chiral components, according to,
\begin{eqnarray}
F \,\psi_{(j_1,j_2)\oplus (j_2,j_1)}&=&c_{(j_1,j_2)}\psi_{ \left(j_1,j_2 \right)\oplus (j_2,j_1)},\,\,
c_{(j_1,j_2)}=\frac{1}{2}\left(K(K+2)+M^2\right), 
\label{F_Cas}\\
G\, \,\phi^R_ {(j_1,j_2)}&=&r_{(j_1,j_2)}\phi^R_{\left(j_1,j_2 \right)}, \quad G\, \,\phi^L_{ (j_2,j_1)}= r_{(j_2j_1)} \phi^L_{(j_2,j_1)},
\label{G_Cas}\\
 r_{(j_1,j_2)}&=&-r_{(j_2j_1)}=i(K+1)M, \quad  K=j_1+j_2, \quad M=|j_1-j_2|.
\label{G_csts}
\end{eqnarray}
The idea of \cite{We} has been to employ the Casimir invariant $F$ in the construction of momentum independent (static) projectors on the irreducible sectors of
the Lorentz-tensor-spinor  in (\ref{prdct_spcs}) and to explore the consequences.
\vspace{0.1cm}
\noindent
The Lorentz projector, denoted by ${\mathcal P}_F^{(3/2,0)\oplus(0,3/2)}$, that identifies the irreducible ${(3/2,0)\oplus (0,3/2)}$ representation space in (\ref{prdct_spcs}), is constructed from $F$ in (\ref{F_Cas}) as,
\begin{eqnarray}
\mathcal{P}_F^{(3/2,0)\oplus(0,3/2)}&=&  \left( \frac{F-c_{(1,1/2)}}{c_{(3/2,0)}-c_{(1,1/2)}} \right)
\left( \frac{F-c_{(1/2,0)}}{c_{(3/2,0)}-c_{(1/2,0)}} \right),\nonumber\\
 c_{(1/2,0)}&=&\frac{3}{4},\quad c_{(3/2,0)}=\frac{15}{4},\quad  c_{(1,1/2)}=\frac{11}{4}.
 \label{General_L_Proj}
\end{eqnarray}
The equation (\ref{General_L_Proj})
shows that the operator $\mathcal{P}_F^{(3/2,0)\oplus(0,3/2)}$ has the property to nullify any irreducible representation space  which is different from
${(3/2,0)\oplus (0,3/2)}$. Instead, for ${(3/2,0)\oplus (0,3/2)}$, it acts as the identity operator, meaning that $\mathcal{P}_F^{(3/2,0)\oplus(0,3/2)}$
is a projector on this very space. In recalling the notation of the spin-$3/2$ wave function, $\left[\Psi^{(3/2,0)\oplus(0,3/2)}_{\left[ \mu, \nu\right]}\right]_a$, we find
\begin{equation}
\left[ \mathcal{P}^{(3/2,0)\oplus(0,3/2)}_F\right]_{\left[ \mu, \nu \right]}{}^{\left[ \eta ,\rho \right]}\, _{a}\, ^b\left[\Psi^{(3/2,0)\oplus(0,3/2)}_{\left[ \eta ,\rho \right]}\right]_{b}
=\left[\Psi^{(3/2,0)\oplus(0,3/2)}_{\left[ \mu ,\nu\right] }\right]_{a}.
\label{General_L_Proj_1}
\end{equation}
The simplest way to bring in kinematics is to set $\left[\Psi^{(3/2,0)\oplus(0,3/2)}_{\left[ \mu ,\nu\right] }\right]_{a}$ on its mass shell, 
\begin{equation}
p^2\left[ \mathcal{P}^{(3/2,0)\oplus(0,3/2)}_F\right]_{\left[ \mu, \nu \right]}{}^{\left[ \eta ,\rho \right]}{}_{a}{}^b\left[\Psi^{(3/2,0)\oplus(0,3/2)}_{\left[ \eta ,\rho \right]}\right]_{b}
=m^2\left[\Psi^{(3/2,0)\oplus(0,3/2)}_{\left[ \mu ,\nu\right] }\right]_{a}.
\label{Master32}
\end{equation}
The symmetry of (\ref{Master32}) is the direct product of the group of translations, ${\mathcal T}_{(1+3)}$, in $(1+3)$ space time, with the spin-Lorentz group, i.e.
${\mathcal T}_{(1+3)}\otimes {\mathcal L}$, some times termed to as the {\tt inhomogeneous spin-Lorentz group}.
In this way, it becomes possible to describe a particle residing in the pure spin-$3/2$ representation space, ${(3/2,0)\oplus (0,3/2)}$,
of the $so(1,3)$ algebra of the spin-Lorentz group, by a Lorentz-tensor-Dirac-spinor wave function, and by means of the one sole  second-order
differential equation in (\ref{Master32}). It has been shown in \cite{We} that within this scheme, 
one reproduces the precise electromagnetic  multipole moments obtained within the canonical Joos-Weinberg method, where the field is described by an eight component ``bi-vector'' \cite{DelgadoAcosta:2012yc}. It has  been furthermore checked that also the properties of the remaining two sectors, known from elsewhere, are correctly reproduced, as it should be. Moreover, in \cite{We} the proof that the wave equation in (\ref{Master32}) is free from the Velo-Zwanziger problem has been delivered by demonstrating that  the
wave fronts of its (classical) solutions propagate always causally within an electromagnetic environment. 
The above presentation should leave it clear that the scheme does not restrict to product spaces of the type given in (\ref{prdct_spcs}) but extends to any product spaces containing the pure spin ${(j,0)\oplus (0,j)}$ of interest. In particular, it applies to bases constructed as direct products of four-vectors, or, to general Weyl-Van-der-Waerden tensor-spinors, as used in the Bargmann-Wigner framework \cite{BWpost}. Moreover, the scheme also extends to high- spins carried by two-spin valued representation spaces of the type, ${(1/2,j-1/2)\oplus (j-1/2,1/2)}$, \cite{IJMPE}, in which case the second order wave equation is not obtained from the mass shell-condition alone but from combining it by another covariant  projector, ${\mathcal P}_{{\mathcal W}^2}^{(j,m)}=(-{\mathcal W}^2/m^2 -j(j-1)p^2/m^2)/(2j)$, which projects besides on the mass shell, also on the highest of the two spin  degrees of freedom. Here, ${\mathcal  W}^2$ is the squared Pauli-Lubanski operator, the second Casimir invariant of the algebra of {\tt inhomogeneous spin-Lorentz group}.
Such a projector technique has first been employed by Aurilia and Umezawa in \cite{Umezawa} at the free particle level, and independently, almost 40 years later by \cite{Napsuciale:2006wr} at the interacting level. In the latter work the second order differential equation following from ${\mathcal P}_{{\mathcal W}^2}^{(j,m)}$ has been extended by the  most general
terms allowed by relativity and containing $\partial_\mu$, $\partial_\nu $ commutators. These terms,  identically vanishing at the free particle level, provide upon gauging essential contributions proportional to the electromagnetic field strength tensor, $F_{\mu\nu}$ and guarantee that the resulting wave equation is  free from the Velo-Zwanziger problem for a $g$ factor taking  the value of $g=2$. 
However, the solutions to second order equations, be them to the representation unspecific  Klein-Gordon--, and Proca types, or to the representation specific equation  in (\ref{Master32}),  describe arbitrary mixing  of  spin-$j$ particles of opposite parities. This arbitrariness can be  removed by the aid of the second Casimir invariant $G$ in (\ref{Cas2}), the subject of the next subsection.    

\subsection{ Separating  left from right chiral degrees of freedom and identifying the parity states}
\vspace{0.1cm}
In parallel to (\ref{General_L_Proj}), also the $G$ invariant of the spin-Lorentz group algebra can be employed in the construction of projector 
operators within any $(j_1,j_2)\oplus (j_2,j_1)$ representations space. Such operators, here denoted by $\mathcal{P}_G^{(j_1,j_2)}$ and $\mathcal{P}_G^{(j_2,j_1)}$, respectively,
have the property to separate the reps under consideration into left (L)- and right (R)-handed degrees of freedom according to, 
\begin{eqnarray}
\mathcal{P}_G^{(j_1,j_2)}=\frac{1}{2}\frac{G+r_{(j_1,j_2)}}{r_{(j_1,j_2)}}, &\quad&
\mathcal{P}_G^{(j_1,j_2)}=-\frac{1}{2}\frac{G-r_{(j_1,j_2)}}{r_{(j_1,j_2)}}.
\label{PR_properties}
\end{eqnarray}
Therefore, from  (\ref{PR_properties}) one observes  that the projectors $\mathcal{P}_G^{(j_1,j_2)}$ and $\mathcal{P}_G^{(j_2,j_2)}$  decompose 
the $2(2j_1+1)(2j_2 +1)$ degrees of freedom residing in  $(j_1,j_2)\oplus (j_2,j_1)$ into the two independent right-handed,
 $(j_1,j_2)$, and  left-handed, $(j_2,j_1)$, irreducible sectors, each having half, i.e. $(2j_1+1)(2j_2+1)$, of the independent degrees of freedom of the initial representation space. Then the two equations,
\begin{equation}
\frac{p^2}{m^2}{\mathcal P}_G^{(j_1,j_2)}{\mathcal P}_F^{(j_1,j_2)\oplus(j_2,j_1)}\phi^R_{\left(j_1,j_2\right)}=\phi^R_{\left(j_1,j_2\right)},\quad
\frac{p^2}{m^2}{\mathcal P}_G^{(j_2,j_1)}{\mathcal P}_F^{(j_1,j_2)\oplus(j_2,j_1)}\phi^L_{\left(j_2,j_1\right)}=\phi^L_{\left(j_2,j_1\right)},
\label{chiral}
\end{equation}
with $j_2=0$, uniquely fix the chiral components of the representation space under consideration. The chiral components are defined by  the symmetric and anti-symmetric combinations of wave functions of
positive and negative parity states, a reason for which the parity states spanning the representation space of interest  can be recovered without ambiguities from the chiral states. Specifically for the Dirac spinor, $j_1=1/2, j_2=0$, where the spin-Lorentz group generators are $1/2\sigma_{\mu\nu}$, the $G$ invariant calculates as, $-3/4i\gamma^5$, thus defining the corresponding projector operators as ${\mathcal P}^{(1/2,0)}_G=(1-\gamma_5)/2$, and ${\mathcal P}_G^{(0,1/2)}=(1+\gamma_5)/2$, leading to the respective Dirac's chiral spinors, $(u-v)/2$, and $(u+v)/2$.

\section{ Summary and conclusions}
We suggested a universal recipe for calculating the states of particles with any high spin-$j$, covariantly transforming according to $(j,0)\oplus (0,j)$, and in terms of Lorentz-tensor-Dirac-spinors (\ref{General_L_Proj_1}), or, Weyl--Van--der-Waerden tensor-spinors \cite{BWpost}, thus facilitating vertex constructions by simple contractions of indexes, and avoiding rectangular-matrix insertions. 
To avoid the problems of the high-order differential equations,
we, similarly to the Fierz-Pauli method, approximated the kinematics through the mass-shell condition alone, and kept the reference frame specification  of the reps (encoded by the boost) solely  at the wave-function level (generated in any frame by construction),
while dropping it from  the wave equation.  Boost incorporation  into the wave equations necessarily rises their order. 
The wave equations in our method are derivable from a Lagrangian. To be specific, within the spin-Lorentz group projector method  pure spin-$3/2$
transforming in the totally symmetric Lorentz tensor of second rank with Dirac spinor components, $\Psi_{\mu,\nu}^{(3/2,0)\oplus(0,3/2)}$, has been described by means of the following Lagrangian \cite{We},

\begin{eqnarray}
{\mathcal L}^{\left(3/2,0\right)\oplus (0,3/2) }_\textup{free}&=&{\Big(}\partial^\mu
[\overline{\Psi}^{\left(3/2,0\right)\oplus (0,3/2)}]^A{\Big)}
[\Gamma_{\mu\nu}^{\left(3/2,0\right)\oplus (0,3/2)}]_{AB}\partial^\nu[\psi^{\left(3/2,0\right)\oplus (0,3/2)}]^B\nonumber\\
&-&m^2[\overline{\Psi}^{\left(3/2,0\right)\oplus (0,3/2)}]^A[\Psi^{\left(3/2,0\right)\oplus(0,3/2)}]_A,
\quad A=\left[\mu\nu\right],\quad B=\left[ \gamma\delta\right],
\label{Lagr}
\end{eqnarray}
with the $\Gamma$-tensor being defined as,
\begin{eqnarray}
~\left[
\Gamma^{\left(3/2, 0\right)\oplus(0,3/2)}{}^\mu{}{}_\nu
\right]^{\left[ \alpha \beta\right]  }{}{}_{ \left[\gamma\delta\right]}&=&4\left[\mathcal{P}^{\left(3/2,0\right)\oplus(0,3/2)}_F\right]
^{\left[\alpha\beta\right]\left[{ \sigma} \mu\right]}\
\left[\mathcal{P}^{\left(3/2,0\right)\oplus (0,3/2)}_F\right]_{\left[\sigma \nu\right]\left[\gamma\delta \right]}\nonumber\\
=\frac{1}{2}{\Big(}
{\sigma}^{\alpha\beta}{ \sigma}^{\sigma\mu}{\mathbf \sigma}_{\sigma\nu}{\sigma}_{\gamma\delta}+
{ \sigma}^{\sigma\mu}{\sigma}^{\alpha\beta}{\sigma}_{\gamma\delta}{\sigma}_{\sigma\nu}
&-&3{ \sigma}^{\alpha\beta}{\sigma}^{\sigma\mu}{ \sigma}_{\gamma\delta}
{\sigma}_{\sigma\nu}-3{\sigma}^{\sigma\mu}{ \sigma}^{\alpha\beta}{\sigma}_{\sigma\nu}
{\sigma}_{\gamma\delta}
{\Big)}
+4{\sigma}^{\sigma\mu}{ \sigma}^{\alpha\beta}{\sigma}_{\sigma\nu}{ \sigma}_{\gamma\delta},
\label{gamma-s32}\nonumber\\
\end{eqnarray}
where $\sigma_{\mu\nu}$ stands for the standard totally anti-symmetric Dirac tensor of second rank.
Along this line, Lagrangians for any spin can be constructed on the cost of increasing the number of the Lorentz indexes. 
We tested this approach in the evaluation
of Compton scattering processes in \cite{We}, \cite{IJMPE} and found that, compared to the Joos-Weinberg bi-vector basis, it notably speeds up the calculations by allowing employment of the FeynCalc software. 
 
\vspace{0.1cm}
\noindent
{\bf Acknowledgments:} We thank the Editors for their kind invitation to contribute to this volume of historical relevance, in honor of the important impact of the DPyC on physics in Mexico.

\end{document}